\documentclass[12pt]{article}
\def\today{January 20, 2004}

\usepackage{amsmath,amssymb}
\usepackage{cite,psfig}

\newcommand{\lspan}{\operatorname{span}}

\newcommand{\up}[1]{^{(#1)}}

\newcommand{\ZZ}{{\mathbb Z}}
\newcommand{\cM}{{\mathcal M}}

\newcommand{\cP}{{\mathcal P}}

\newcommand{\fD}{{\mathfrak D}}

\newcommand{\pa}{\partial}

\newcommand{\myspan}{\operatorname{span}}

\newcommand{\e}{{\rm e}}
\newcounter{mylc}
\renewcommand{\themylc}{\roman{mylc}}

\newtheorem{proposition}{Proposition}
\newtheorem{theorem}{Theorem}

\newtheorem{corollary}{Corollary}
\newtheorem{lemma}{Lemma}
\newcommand{\qed}{QED}
\begin{document}
\title{Quasi-Exact Solvability and the direct approach to invariant subspaces}
\author{D. G\'omez-Ullate\thanks{CRM, Montreal, Canada. E-mail:
ullate@crm.umontreal.ca}, N. Kamran,\thanks{McGill University,
Montreal, Canada. . E-mail: nkamran@math.mcgill.ca} and R.
Milson\thanks{Dalhouise University, Halifax, Canada. E-mail:
milson@mathstat.dal.ca} }

\date{\today}
%
%
\maketitle
\begin{abstract}
We propose a more direct approach to constructing differential
operators that preserve polynomial subspaces than the one based
on considering elements of the enveloping algebra of
$\mathfrak{sl}(2)$. This approach is used here to construct new
exactly solvable and quasi-exactly solvable quantum Hamiltonians
on the line which are not Lie-algebraic. It is also applied to
generate potentials with multiple algebraic sectors. We discuss
two illustrative examples of these two applications: an
interesting generalization of the Lam\'e potential which posses
four algebraic sectors, and a quasi-exactly solvable deformation
of the Morse potential which is not Lie-algebraic.
\end{abstract}

\section{Introduction}

Our purpose in this paper is to show that the property of
quasi-exact solvability for Schr\"odinger operators admits an
effective formulation which goes beyond the Lie algebraic context
of hidden symmetry algebras, and to show explicitly that several
classes of physically relevant potentials are quasi-exactly
solvable in this more general sense. These include novel
generalizations of the Lam\'e and Morse potentials.

To put our results in context, we begin by briefly recalling the
general definition of quasi-exact solvability, as well as the Lie
algebraic formulation which applies to some particular classes of
invariant subspaces.

Consider a one-dimensional Schr\"odinger operator
\begin{equation}\label{schr}
 H = -\partial_{xx} + U(x),
\end{equation}
which is assumed to be essentially self-adjoint on a Hilbert space
${\mathcal H}$ of wave functions. We say that $H$ is {\em
quasi-exactly solvable} (QES) \cite{Tu88,KO90} if it leaves
invariant a non-trivial finite-dimensional subspace $\mathcal M
\subset {\mathcal H}$
\[ H \mathcal M\subset \mathcal M\]
where $\mathcal M$ is  \[\mathcal M=
\myspan\{\phi_1,\dots,\phi_n\},\quad \phi_i\in \mathcal H.\] In
this case the restriction of $H$ to $\mathcal M$ is represented
by a finite dimensional matrix, and a finite portion of the
spectrum of $H$ can be obtained algebraically by diagonalizing
this matrix. If $H$ actually preserves a complete flag of
subspaces
\[\cM_1\subset \cM_2\subset\dots\subset \cM_n\subset\dots\]
then it is said to be {\em exactly solvable} (ES). Given a
certain potential $U(x)$  it is in general very difficult to
determine whether $H$ preserves some non-trivial invariant
subspace $\mathcal M$. One generally proceeds in the opposite
direction, and starting from a known finite-dimensional subspace
$\cal M$ one attempts to determine all Hamiltonians $H$ which are
known {\em a priori} to leave $\cal M$ invariant. In most
applications  $\cal M$ is assumed to be of the form
\begin{eqnarray}
&{\cal M}_n &= \mu(x) \cP_n(z(x)),\\
&\cP_n&= \langle 1,z,z^2,\dots,z^n \rangle,
\end{eqnarray}
where $\mu$ is some smooth non-vanishing multiplier, so that the
first problem is to construct second-order differential operators
which leave the polynomial module $\cP_n$ invariant. In this
context, Burnside's theorem ensures that every operator which
leaves $\cP_n$ invariant is an element of the enveloping algebra
of the Lie algebra $\mathfrak{sl}(2)$ in the following standard
representation
\begin{equation}\label{2Js}
J^+_n=z^2\partial_z-n z\,,\qquad J^0_n=z\partial_z-\frac
n2\,,\qquad J^-_n=\partial_z\,.
\end{equation}
This approach is known as the Lie algebraic approach and
$\mathfrak{sl}(2)$ is said to be a hidden symmetry algebra of the
Hamiltonian. The Lie-algebraic approach has been extended to
higher dimensional problems: some examples of Lie-algebraic
Hamiltonians in two dimensions are given in \cite{GKO94}, while
the quantum $N$-body Calogero-Moser model has been shown to have
hidden symmetry algebra $\mathfrak{sl}(N+1)$ \cite{Tu94}, and
many-body problems with elliptic interaction were shown to be
Lie-algebraic in \cite{GGR01}.

In the last twenty five years, most of the papers dealing with
quasi-exact solvability have made use of the Lie-algebraic
approach, using irreducible representations of finite-dimensional
Lie algebras by first-order differential operators as building
blocks for constructing differential operators with invariant
subspaces. Despite the success of this method, there remain some
natural questions which are not amenable to this Lie-algebraic
approach. For example, one would like to be able to construct the
most general differential operator that will map a polynomial
module $\cP_n$ into a proper subspace $\cP_{n-k}\subset \cP_n$ or
to know what are the operators that preserve {\em monomial
subspaces}, that is subspaces of $\cP_n$ which admit a monomial
basis, but where all degrees up to $n$ need not be present (see
definition \eqref{gapdef} in Section \ref{gap}). It is the latter
question that we consider in Section 2 of this paper, where we
obtain a general structure theorem for the space of linear
differential operators preserving a general monomial subspace.
Much of the contents of this Section can already be found in a
paper by Post and Turbiner in \cite{PT95}. Sections 3 and 4 form
the heart of our paper. Section 3 consists in an explicit
application of the results of Section 2 to Hamiltonians with
multiple algebraic sectors viewed from the perspective of
monomial subspaces/ This Section also includes an intriguing
generalization of the Lam\'e equation which is shown to admit
four algebraic sectors instead of the double algebraic sector
admitted by the classical Lam\'e equation. In Section 4 we
discuss a modification of the Morse potential which is shown to
preserve a proper monomial subspace of $\cP_n$. This class of
potentials cannot be obtained in the Lie-algebraic approach and
therefore they are not present in the classification of
Lie-algebraic potentials on the line \cite{GKO}.

\section{The direct approach to invariant monomial subspaces}\label{gap}

In this Section we shall introduce the concept of a {\em monomial
subspace } and we will characterize all differential operators of
order $r$ that preserve a given monomial subspace. This Section
follows essentially the algebraic approach introduced in
\cite{PT95}.

For a finite subset $I\subset \ZZ$ let us define a {\em monomial
subspace } $\cP_I$ as
\begin{equation}\label{gapdef}
\cP_I=\lspan\{ z^k : k\in I\},
\end{equation}
and let $N=|I|$ be the dimension of $\cP_I$. A non-trivial
differential operator $T^d_r$ of order $r$ will be said to be of
degree $d \in \ZZ$ if and only if
\begin{equation}
T^d_r [x^j]= c_j \,x^{j+d}\quad \forall j\in \ZZ.
\end{equation}
Let us denote by $\fD_r(\cP_I)$ the space of differential
operators  of degree $r$ with analytic coefficients that preserve
$\cP_I$.
\begin{proposition}
Every linear transformation of  $\cP_I$ can be represented by an
element of $\fD_{N-1}(\cP_I)$.
\end{proposition}
{\em Proof.}
 A basis of the dual space is given by the following
$(N-1)$-th order operators
\begin{equation}\label{duals}
 L_k = z^{-k} \prod_{\substack{j\in I\\j\ne k}}  \frac{z\partial_z -
      j}{k-j},\quad k\in   I,
\end{equation}
which clearly satisfy $L_k [z^j] = \delta_k^j,\; \forall j,k \in
I$. Therefore, every element of $\text{End}(\cP_I)$ can be
represented by an operator of the form
$$ T_{N-1}=\sum_{j,k \in I}a_{jk} z^j L_k \in \fD_{N-1}(\cP_I)$$\qed

In fact, these operator duals can be constructed for more general
subspaces of functions (not just polynomials) \cite{KMO}, but the
order of the differential operators grows with the dimension of
the space. We are interested in constructing differential
operators that preserve $\cP_I$ of order $r<N-1$. Let us denote by
$\fD^d_r(\cP_I) \subseteq \fD_r(\cP_I)$ the space of operators of
order $r$ and degree $d$ that preserve $\cP_I$. In order to
characterize $\fD_r(\cP_I)$ we can concentrate in operators
having a fixed degree, since the following Proposition holds.
\begin{proposition}
\begin{equation}
\fD_r(\cP_I)= \bigoplus_d \fD^d_r(\cP_I)
\end{equation}
\end{proposition}
\emph{proof} Any operator $T \in \fD_r(\cP_I)$ can be decomposed
uniquely as a sum $T=\sum_d T^{(d)}$ of operators of fixed degree.
Then it suffices to prove  that if $T$ preserves $\cP_I$ then each
$T^{(d)}$ also preserves $\cP_I$. Since $T \cP_I \subset \cP_I$,
then
$$ T[z^k] = \sum_{j\in I} a_{jk} z^j,\quad k\in I,$$
and it follows that
$$T^{(d)}[z^k] =
  \begin{cases}
    a_{d+k,k}\, z^{d+k}, \quad & \text{if } d+k\in I;\\
    0 &\text{otherwise.}
  \end{cases}$$
which proves that $T^{(d)} \cP_I \subset \cP_I$.\qed

Let us define the {\em primitive index sets}
\begin{equation}\label{prindex}
I(a,b,c) = \{a + kc: 0\leq k \leq b\}
\end{equation}
 and the {\em primitive monomial subspaces} $\cP_{a,b,c}$ as
\begin{equation}\label{elemgapdef}
\cP_{a,b,c}= \cP_{I(a,b,c)} = \lspan\{z^{a+kc}: 0\leq k \leq b\}
\end{equation}
where $a$ is the smallest exponent, $c$ is the size of the gap and
$b+1$ is the number of elements. In fact,
$$\cP_{a,b,c} = z^a \cP_b (z^c),$$
where $\cP_n(z)= \langle 1,z,\dots,z^n \rangle$ denotes the full
 polynomial module (no gaps). Every monomial subspace can be trivially
decomposed as a direct sum of singleton primitive monomial
subspaces
$$\cP_I = \bigoplus_{a\in I} \cP_{a,0,0},$$ which corresponds
to representing $I$ as the union of $N$ singletons. The
differential operators \eqref{duals} of order $N-1$ are
constructed to annihilate $N-1$ different exponents. However, it
is precisely when the index set $I$ can be expressed as the union
of a smaller number of primitive index sets \eqref{prindex}, that
lower order operators preserving $\mathcal P_I$ can exist. This
consideration leads to the following key theorem:

\begin{theorem}\label{gapthm}
If the set $I$ can be expressed as the union of a finite number of
primitive index sets of the form
\begin{equation}
    \label{Idecomp}
    I=\bigcup_{i=1}^r I(a_i,b_i,d),
  \end{equation}
then there exist  operators $T_r^{(\pm d)}$ of order $r$ and
degree $\pm d$ that preserve $\cP_I$.  Conversely, if $T_r^{(d)}$
is an operator of order $r<N$ and degree $d>0$ that preserves a
given monomial subspace $\cP_I$, then there exists a $T_r^{(-d)}$
which also preserves $\cP_I$ and a vector-space decomposition of
$\cP_I$ into primitive monomial subspaces
  \[\cP_I = \sum_{i=1}^r \cP_{a_i,b_i,d}.\]
\end{theorem}

\emph{Proof.}
  Suppose that \eqref{Idecomp} holds and consider the following $r$-th order operators
  \begin{align}\label{Td}
    T_r\up{-d} &= z^{-d} \prod_{i=1}^r (z\partial_z - a_i)\\
    T_r\up{d} &= z^{d} \prod_{i=1}^r (z\partial_z - (a_i+d b_i))
  \end{align}
  The lowering operator maps $z^{a_i+kd}$ to
  $z^{a_i+(k-1)d}$ and annihilates $z^{a_i}$ for all $i=1,\ldots,r$.
  Similarly the raising operator maps $z^{a_i+kd}$ to
  $z^{a_i+(k+1)d}$ annihilates
  $z^{a_i+db_i}$ for all $i$. It is clear then that \[ T_r\up{\pm d} \cP_I \subset \cP_I.\]
To prove the converse we assume that $T\up{d}$ is an
  operator of degree $d>0$ and order $r$ that preserves a given
  $\cP_I$. The operator $z^{-d} T\up{d}$ has degree 0, and hence has
  a unique factorization of the form
  $$z^{-d}\, T\up{d} = c \prod_{i=1}^r (z\partial_z - k_i),$$
  up to a multiplicative constant $c$.
  For each $i$, let $b_i$ to be the largest possible natural number
  such that  $I(a_i, b_i, d)\subset I,$  where $a_i = k_i-d b_i$.

  If \eqref{Idecomp} did not hold, suppose that $\alpha$ is the largest
  element of $I$ that is not in any of the $I(a_i,b_i,d)$.  Since
  $\alpha \neq k_i$ for all $i$, $T\up{d}$ does not
  annihilate $z^\alpha$ and by assumption $\alpha+d\in I$. But since $\alpha$ was the largest element
  in $I$  which does not belong to any of the $I(a_i,b_i,d)$, we
  must have $\alpha+d\in I(a_i,b_i,d)$ for some $i$, but then by
  the way $b_i$ was chosen,  $\alpha \in I$ as well, which is a contradiction. \qed

\subsection{Second order operators}

The case of second order operators is particularly important for
the applications in quantum mechanics, and we shall concentrate to
their study in this Section. From Theorem \ref{gapthm} we know
that if a given monomial subspace $\cP_I$ is invariant under
$T_2\up{d}$, it is also invariant under $T_2\up{-d}$, and of
course it always happens that $T_2\up{0} \in \fD_2(\cP_I)$ for any
monomial subspace $\cP_I$. The natural question is now to explore
whether a given monomial subspace can be preserved by several
operators of different degrees.

\begin{theorem}\label{excepthm}
Suppose that a monomial subspace $\cP_I$ of dimension $|I|>3$ is
invariant under two second order operators $T\up{d_1}_2$ and
$T\up{d_2}_2$ of degrees $d_1,d_2 \in \mathbb Z^+$ with $d_1>d_2$
coprimes. Then $d_1=2$, $d_2=1$ and $\mathcal P_I$ must be one of
the following $3$ modules:
\begin{enumerate}
\item[i)] $\cP_n= \langle 1,z,z^2,\dots,z^n \rangle$
\item[ii)] $\tilde{\cP}_n= \langle 1,z,z^2,\dots,z^{n-2},z^n
\rangle$
\item[iii)] $\hat{\cP}_n= \langle 1,z^2,\dots,z^n \rangle$
\end{enumerate}
\end{theorem}

\emph{Proof.} Let $I = \{i_1,\dots,i_N\}$ with
$i_1<i_2<\dots<i_{N-1}<i_{N}$. Both $T_2^{(d_1)}$ and $
T_2^{(d_2)}$ must annihilate $z^{i_N}$, so they must be of the
form
$$
\begin{aligned}
T_2^{(d_1)} = z^{d_1}(z\pa_z - i_N)(z\pa_z - a_1)\\
T_2^{(d_2)} = z^{d_2}(z\pa_z - i_N)(z\pa_z - a_2)
\end{aligned}
$$ with $a_1$ and $a_2$ still to be specified. The difference
$i_N-i_{N-1}$ must be equal to either $d_1$ or $d_2$ because
otherwise both $T_2^{(d_1)}$ and $ T_2^{(d_2)}$ must annihilate
$z^{i_{N-1}}$ too, fixing $a_1=a_2=i_{N-1}$. But then either
$T_2^{(d_1)}[z^{i_{N-2}}]$ or $ T_2^{(d_2)}[z^{i_{N-2}}]$ will be
outside $\mathcal P_I$. So two possibilities remain
\begin{enumerate}
\item $i_N-i_{N-1} = d_1$

In this case $ T_2^{(d_2)}$ must annihilate $z^{i_{N-1}}$ so
$a_2=i_{N-1}$, but this forces $i_{N-1}-i_{N-2} = d_2$ since
otherwise $ T_2^{(d_2)}[z^{i_{N-2}}] \notin \mathcal P_I$. It
follows that $ T_2^{(d_1)}[z^{i_{N-2}}]=0$  fixing $a_1=i_{N-2}$.
But now neither $T_2^{(d_2)}$ nor $T_2^{(d_1)}$ can annihilate
$z^{i_{N-3}}$, so $i_{N-2}-i_{N-3} = d_2$ and $i_{N-1}-i_{N-3} =
d_1$. This implies $d_1=2d_2$ and since they are coprimes  $d_1=2$
and $d_2=1$. The preserved module is then $\tilde{\mathcal P}_n$.

\item $i_N-i_{N-1} = d_2$

In this case $T_2^{(d_1)}[z^{i_{N-1}}]=0$ so $a_1=i_{N-1}$. The
difference $i_{N-1} - i_{N-2}$ cannot be $d_1$ because otherwise
$T_2^{(d_2)}$ would need to annihilate $z^{i_{N-2}}$, and then
either $T_2^{(d_1)}[z^{i_{N-3}}]$ or $T_2^{(d_2)}[z^{i_{N-3}}]
\notin \mathcal P_I$. So $i_{N-1} - i_{N-2}=d_2$ but then
$T_2^{(d_1)}[z^{i_{N-2}}]\propto z^{i_{N}}$ and this can only
happen if $d_1=2d_2$. Iterating these arguments it is easy to see
that all differences between successive indices must be equal to
$d_2$ except for $i_2-i_1$ which could be either $d_1$ (which
leads to $\hat{\mathcal P}_n$) or $d_2$, which leads the ordinary
polynomial module $\mathcal P_n$.\qed
\end{enumerate}

\begin{corollary}
No monomial subspace $\cP_I$ is invariant under three second order
operators of relatively prime degrees $d_1,d_2,d_3 \in \mathbb
Z^+$.
\end{corollary}
\emph{Proof.} It is clear that a third operator $T_2^{(d_3)}$ with
an odd $d_3>2$ will not preserve any of the modules of the
previous Theorem.

\begin{corollary}\label{cor_dim}
The spaces of  second order differential operators that leave
$\cP_n$, $\hat \cP_n$ and $\tilde \cP_n$ invariant have the
following dimensions:
$$
\dim \fD_2(\cP_n)=9, \quad \dim \fD_2(\hat\cP_n)=\dim
\fD_2(\tilde \cP_n) =7.
$$
\end{corollary}

\emph{Proof.} For the case of $\mathcal{P}_n$, the dimensions of
the corresponding operator spaces $\fD_2^k(\cP_n)$ are:
\begin{equation}
\dim \fD_2^0(\cP_n)=3\,,\quad \dim \fD_2^{\pm 1}(\cP_n)=2\,,\quad
\dim \fD_2^{\pm 2}(\cP_n)=1.
\end{equation}
This follows from noting that the operators in question can be
expressed in the form
\begin{eqnarray*}
  T^{(0)}&=& p(z\partial_z),\quad p(x)\in \cP_2(x),\\
  T^{(1)}&=& z (a_1 z\partial_z-b_1) (z\partial_z-n),\\
  T^{(-1)}&=& z^{-1} (a_2 z\partial_z-b_2)  (z\partial_z),\\
  T^{(-2)} &=& z^{-2}(z\partial_z-1)(z\partial_z),\\
  T^{(2)} &=& z^2 (z\partial_z-n)(z\partial_z-(n-1)).
\end{eqnarray*}
The situation for the case of $\hat{\mathcal{P}}_n$ is similar,
the difference being that the dimension of $\fD_2^{\pm
1}(\hat\cP_n)$ is only one since the corresponding operators must
have the form
\begin{eqnarray*}
  T^{(1)}&=& a_1 z (z\partial_z-0) (z\partial_z-n),\\
  T^{(-1)}&=& a_2 z^{-1} (z\partial_z-0) (z\partial_z-2).
\end{eqnarray*}
Thus the dimension of $\fD_2(\hat\cP_n)$ is seven. \qed
 In order to apply this construction to Quantum
Mechanics, we need to relate an operator $T \in \fD_2(\cP_I)$
with a Schr\"odinger operator \eqref{schr}. This is always
possible in one dimension by virtue of the following Lemma:
\begin{lemma}\label{lematransf}
Every second order operator $T$ of the form
\begin{equation}\label{eq:Pzz}
T=P(z)\partial_{zz} + \big(Q(z)+\frac{1}2 P'(z)\big)\partial_z +
R(z),
\end{equation}
with $P(z)>0$ can be transformed into Schr\"odinger form
\eqref{schr} by a gauge transformation and a change of variables:
\begin{equation}\label{gaugetransf}
 H (x) = -\mu(z) T(z) \mu^{-1}(z)\Big|_{z=\xi^{-1}(x)},
\end{equation}
where
\begin{eqnarray}
  \label{eq:physvar}
&  x &= \xi(z) = \int^{z} \frac{d \zeta}{\sqrt{P(\zeta)}};\\
  \label{eq:gaugeform}
 &  \mu(z) &=\exp  \int^{z}
  \frac{Q(\zeta)}{2P(\zeta)}d\zeta,
\end{eqnarray}
The resulting potential $U(x)$ in the physical variable $x$ is
given by
\begin{equation}
  \label{eq:potential}
  U(x) = R(z) - \frac{1}2 Q'(z) + \frac{Q(z) (P'(z) - Q(z))}{4 P(z)}
  \Bigg|_{z=\xi^{-1}(x)}.
\end{equation}

\end{lemma}

\section{Quantum Hamiltonians with multiple algebraic
sectors}\label{multiple}

We are familiar with the fact that Schr\"odinger operators with
even potentials will have eigenfunctions with well defined
parity, and therefore the odd sector can be treated independently
from the even sector. Let us look at the phenomenon of multiple
algebraic sectors from the point of view of monomial subspaces.
There are two mechanisms that can produce a multiple algebraic
sector. The first one comes from the fact that a second order
operator can preserve two different polynomial modules.  From
Theorem \ref{gapthm} we know that if
$\fD_2\up{d}(\cP_I)\neq\emptyset$, then $\cP_I$ can be decomposed
as
\begin{equation}\label{multimod}
\cP_I = \cP_{a_1,n_1,d}+\cP_{a_2,n_2,d}.
\end{equation}
If in addition $\fD_2\up{d_2} (\cP_I)=\emptyset$ for every other
$d_2 \neq d, d_2\in \mathbb Z^+$ then
\begin{eqnarray}
\fD_2\up{d} (\cP_I) \subset \fD_2\up{d} (\cP_{a_1,n_1,d}),\\
\fD_2\up{d} (\cP_I) \subset \fD_2\up{d} (\cP_{a_2,n_2,d}),
\end{eqnarray}
and both spaces are invariant separately. We can take $a_1=0$
without loss of generality and let $p=a_2/d, p\in \mathbb R$. In
the new variable $y = z^d$ the two invariant polynomial modules
are
\begin{equation}\label{m1m2}
\cM_1= \cP_{n_1}(y),\quad \cM_2 = y^p \cP_{n_2}(y),
\end{equation}
and each of them is independently preserved.
\begin{proposition}\label{prop:m1m2}
The most general second order differential operator that
preserves {\em both} $\cM_1$ {\em and} $\cM_2$ in \eqref{m1m2} is
a linear combination of the following five operators
\begin{eqnarray}\label{t+1}
  &T\up{+1}_2 &= y (y\partial_y - n_1)(y\partial_y - (p+n_2)) \\
 & &= y^3\partial_{yy} +  (1-p-n_1-n_2) y^2 \partial_y + n_1 (n_2+p) y, \nonumber\\
 & T\up{0}_2 &=y^2\partial_{yy}, \label{t0}\\
& T\up{-1}_2 &= y^{-1}(y\partial_y-0)(y\partial_y - p)\label{t-1}\\
& &= y \partial_{yy} +  (1-p) \partial_y, \nonumber\\
  &T\up{0}_1 &=y\partial_y,\label{t10}\\
  &T\up{0}_0 &=1.\label{t00}
\end{eqnarray}
\end{proposition}
These operators form a sub-family of the well known nine parameter
family of Lie-algebraic second order operators that would
preserve the polynomial module $\cM_1$.

A second source of multiple algebraic sectors comes from noting
that in Lemma \ref{lematransf} the correspondence between the
algebraic operator $T$ and the Schr\"odinger operator $H$ is at
least two to one:
\begin{proposition}\label{prop:T1T2}
The operators $T_1$ and $T_2$ given by
\begin{eqnarray}
&T_1 &=P(z)\partial_{zz} + \big(Q(z)+\frac{1}2
P'(z)\big)\partial_z
+ R(z),\\
&T_2 &= P(z)\partial_{zz} + \big(-Q(z)+\frac{3}2
P'(z)\big)\partial_z + R(z)-Q'(z) +\frac{1}2 P''(z)
\end{eqnarray}
are both equivalent under the transformation
\eqref{gaugetransf}--\eqref{eq:gaugeform} to the same
Schr\"odinger operator $H=-\partial_{xx}+U(x)$ with potential
$U(x)$ given by \eqref{eq:potential}.
\end{proposition}
Therefore if both $T_1$ and $T_2$ preserve different finite
dimensional subspaces, then the Schr\"odinger operator will have
a double algebraic sector.

The two mechanisms described above can be combined to yield
potentials with quadruple algebraic sectors. This will be the
case if operators $T_1$ and $T_2$ in Proposition \ref{prop:T1T2}
are both a linear combination of the operators of Proposition
\ref{prop:m1m2}. These ideas lead to a generalization of the
Lam\'e potential that posses a quadruple algebraic sector.

\subsection{The generalized Lam\'e potential}

\begin{figure}[htbp]
  \begin{center}
    \noindent\psfig{figure=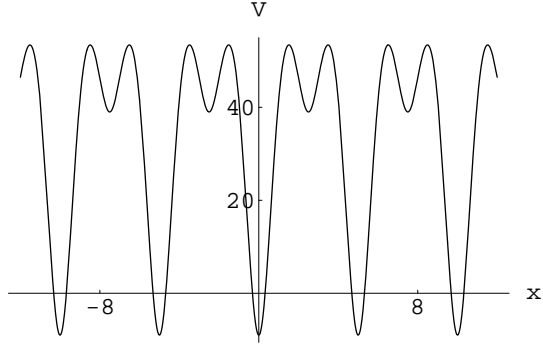,width=3in}
    \caption{ Generalized Lam\'e potential \eqref{potLame} for $m=11/10$, $l=9$ and $j=6$ }
    \label{fig:pot}
  \end{center}
\end{figure}

Let us consider the following linear combination of the operators
(\ref{t+1})-(\ref{t00}):
\begin{equation}
T= 4 k T_2\up{+1} + 4(k+1) T_2\up{0} + 4 T_2\up{-1} +4 b_1
T_1\up{0}
\end{equation}
which can be written in the form \eqref{eq:Pzz} with
\begin{eqnarray}
&P(z)&= 4z(z+1)(kz+1),\\
&Q(z) &= -2k(1+2 n_1 + 2 n_2 + 2p) z^2 + 4(b_1-k-1)z +2-4p,\\
&R(z) &= 4 k n_1 (n_2+p) z.
\end{eqnarray}
The cubic $P(z)$ determines the change of variables
\eqref{eq:physvar} and we must consider only the interval in $z$
such that $P(z)>0$ for the change of variables to be admissible.
If we take $k<0$ and let $z \in (0,-1/k)$ the change of variables
determined by \eqref{eq:physvar} is
\begin{equation}\label{eq:change}
z = k^{-1} \text{sn}^2(\sqrt{-k}\, x|k^{-1}),
\end{equation}
which maps the real line to the interval (0,-1/k). From equation
\eqref{eq:potential} in Lema \ref{lematransf} we see that the
condition for the potential to be nonsingular is that \mbox{$Q(z)
(P'(z)-Q(z))$} should divide $P(z)$, and since $z \in (0,-1/k)$
this implies that
\[
Q(0)=0\quad \text{or}\quad Q(0)=P'(0),\] and \[ Q(-1/k) = 0\quad
 \text{or} \quad Q(-1/k) = P'(-1/k).
\]
The first of these conditions is satisfied when $p=\pm 1/2$ and
since we can assume, without loss of generality, that $0$ is the
smallest power in the monomial subspace, then we must take
$p=1/2$. The second set of conditions amounts to choosing $b_1$ as
\begin{eqnarray}
&b_1 &= k-n_1-n_2,\label{eq:b1}\\
&b_1 &= 2k-1-n_1-n_2.\label{eq:b2}
\end{eqnarray}
Therefore we can write the potential \eqref{eq:potential} after
performing the change of variables \eqref{eq:change} as
\begin{equation}\label{potLame}
V(z)= ml(l+1) \text{sn}^2(x|m) - j(j+1)\,
\text{sn}^2(\sqrt{m-1}\,x|\textstyle{\frac{1}{1-m}})
\end{equation}
where $m=1-k$ and $j,l\in \mathbb Z$ are related to $n_1$ and
$n_2$ as
\begin{eqnarray}
&l=n_1+n_2+1,\quad &j=n_1-n_2-1\\
&l=n_1+n_2+2,\quad &j=n_2-n_1
\end{eqnarray}
depending on whether we use \eqref{eq:b1} or \eqref{eq:b2}. Note
that the particular cases $j=0$ and $j=-1$ correspond to the
Lam\'e equation, and in those cases the two polynomial modules in
\eqref{m1m2} match with no half-integer gaps. However, for
greater values of $j$ the potential departs from Lam\'e and
develops an interesting shape, as can be seen in Figure
\ref{fig:pot}.
Depending on whether we use \eqref{eq:b1} or \eqref{eq:b2} in the
formula for the gauge factor
 \eqref{eq:gaugeform} we obtain
\begin{eqnarray}
&\mu_1&=(1+z)^{-l/2}=\text{cn}^l(x|m),\\
&\mu_2&=(1+z)^{-l/2}(1+kz)^{1/2}=\text{cn}^l(x|m)\,\text{cn}(\sqrt{-k}\,x|\textstyle{\frac{1}{1-m}}),
\end{eqnarray}
To summarize the results in this section, we have proved that
when $j$ and $l$ are integers of the same parity, the generalized
Lam\'e potential \eqref{potLame} has the following quadruple
algebraic sector:
\begin{eqnarray}
 & \cM\up{1}_1 =\mu_1 \,\cP_{n_1}(z),\qquad &
\cM\up{1}_2=\mu_1\,\cP_{n_2}(z),\\
 & \cM\up{2}_1=\mu_2\,\cP_{\tilde n_1}(z),\qquad & \cM\up{2}_2=\mu_2
\,\cP_{\tilde n_2}(z),
\end{eqnarray}
where
\begin{equation}
\begin{aligned}
 n_1 &=\displaystyle \frac{ l+j}2,\quad & n_2 = \frac{l-j}2-1,\\
 \tilde n_1 &= \displaystyle \frac{ l-j}2-1,\quad & \tilde n_2 = \frac{l+j}2-1,
\end{aligned}
\end{equation}
and $z$ is given by \eqref{eq:change}. The total number of
algebraic eigenfunctions is the sum of the dimensions of the four
algebraic sectors, which equals $2 l+1$. It is quite remarkable
that this number does not depend on $j$. In fact, when $j=0$ and
$l\in \mathbb N$, these eigenfunctions are the well known $2l+1$
Lam\'e polynomials. The period of the potential is $2K$ where
\begin{equation}
K = \int_0^\frac{\pi}2 {d \theta \over \sqrt{1-m^2 \sin^2 \theta}}
\end{equation}
is the elliptic integral of the first kind. There are $l+1$
eigenfunctions which belong to the sectors $\cM\up{1}_{1,2}$ have
the same period as the potential, while the $l$ eigenfuntions
which belong to the sectors $\cM\up{2}_{1,2}$ have period $4 K$.
A detailed study of the properties of these polynomial
eigenfunctions shall be deferred to a forthcoming publication.

Let us note as a conclusion that the generalized Lam\'e potential
could be applied as realistic model of a one-dimensional crystal,
which generalizes the models introduced by Alhassid et al. in
\cite{AGI83}. The generalized Lam\'e potential would describe a
one dimensional crystal with constant lattice spacing $a=2K$ and
two atoms of different atomic number. By changing the values of
$j$ and $l$ the relative strength of the potential wells can be
modified. The eigenvalues of the potential can be calculated
algebraically and they describe the edges of the allowed energy
bands of the crystal. Other possible applications of this
potential include the quantum fluctuations of the inflaton field
in certain cosmological models \cite{GKLS97,FGMR00}.

\section{New Quasi-exactly solvable hamiltonians which are not Lie-algebraic}\label{newQES}

In this section we will apply a similar construction to obtain
quasi-exactly solvable potentials that cannot be obtained as a
quadratic polynomial in the generators of $\mathfrak{sl}(2)$. In
order to construct second order operators that are not
Lie-algebraic we should focus our attention in cases \emph{ii})
and \emph{iii}) of Theorem \ref{excepthm}, the so called
exceptional modules, since the full polynomial module $\cP_n$ is
the carrier space for an irreducible representation of
$\mathfrak{sl}(2)$, and we have from Burnside's theorem that any
operator which leaves $\cP_n$ invariant must be a polynomial in
the generators of $\mathfrak{sl}(2)$. Let us note that the two
exceptional modules are projectively equivalent, since
\begin{equation}
\hat{\cP}_n (z)= z^n \tilde{\cP_n}(1/z),
\end{equation}
and we can therefore restrict our attention to one of them.
\begin{proposition}\label{prop5}
  A differential operator of order two or less preserves
  \begin{equation}\label{excepmodule}
\hat{\cP_n}(z) = \lspan\{1,z^2,\dots,z^n\},
  \end{equation}
 if and only if it is a linear combination of the
  following seven operators:
\begin{eqnarray*}
 & T\up{+2}_2  &= z^4\partial_{zz} + 2(1-n)z^3\partial_z + n(n-1)z^2\;,\\
 &T\up{+1}_2   &= z^3 \partial_{zz} +(1-n) z^2\partial_z\;,\\
 & T\up{0}_2 &= z^2\partial_{zz} \;,\\
  &T\up{-1}_2  & = z\partial_{zz} -\partial_z \;,\\
  &T\up{-2}_2   &=\partial_{zz}-2z^{-1}\partial_z\;,\\
  &T\up{0}_1 &= z\partial_z.\\
  &T\up{0}_0 &= 1.
\end{eqnarray*}
\end{proposition}
\emph{Proof.} The sufficiency follows from the fact that each of
the operators above preserves $\hat P_n$, which can be verified by
direct computation. The necessity is a direct consequence of
Corollary \ref{cor_dim}, and the fact that the above operators are
linearly independent.

Consider the action of $SL(2,\mathbb R)$ on the (projective) line
according to the linear fractional, or M\"obius transformations
\begin{equation}\label{moebius}
w\mapsto z= {\alpha w + \beta \over \gamma w+ \delta}, \quad
\Delta = \alpha \delta - \beta \gamma =1.
\end{equation}
There is an induced action of $SL(2)$ on the space $\cP_n$ of
polynomials of degree at most $n$, given by
\begin{equation}\label{proj_action}
P(z)\mapsto \tilde P(w)= (\gamma w+\delta)^n P\left({\alpha w +
\beta \over \gamma w+ \delta}  \right).
\end{equation}
This irreducible multiplier representation of $SL(2)$ is
isomorphic to the standard representation on homogeneous
polynomials of degree $n$ in two variables. Since $\cP_n$ is
invariant under this transformation, it was used in \cite{GKO94}
to classify the second order operators that preserve $\cP_n$ into
equivalence classes, each of them labeled by a simple canonical
form. On the contrary, monomial subspaces (and in particular the
exceptional module $\hat \cP_n$) are not invariant under the
transformation \eqref{proj_action}. This is clear since $\hat
\cP_n \subset \cP_n$ is a hyperplane of co-dimension 1 in $\cP_n$
which  moves under the projective $SL(2)$ action. In particular,
under the transformation \eqref{proj_action} the exceptional
module $\hat \cP_n (z)$ transforms into
\begin{equation}\label{Mbasis}
\begin{aligned}
&\hat\cP_n(z)\mapsto\cM_n(w) \subset \cP_n(w),\\
&\cM_n(w)=\lspan\left\{(\gamma w + \delta)^{n-k} ( \alpha w +
\beta)^k \,|\,k=0,2,\dots,n\right\}.
\end{aligned}
\end{equation}
It will be convenient to introduce a basis independent
characterization of the space $\cM_n$.
\begin{proposition}\label{prop_basis}
If $\alpha,\gamma \neq 0$ a polynomial $P(w)$ of degree $n$ is in
$\cM_n\subset \cP_n$ if and only if
\begin{equation}\label{gapcondition}
P'(-\beta/\alpha) - n\alpha \gamma P(-\beta/\alpha)=0.
\end{equation}
\end{proposition}
\emph{Proof.} Since the condition is linear, it suffices to prove
that it is satisfied by all the elements in the basis
\eqref{Mbasis} of $\cM_n(w)$. It is clear that all elements with
$k\geq 2$ satisfy $P'(-\beta/\alpha)=P(-\beta/\alpha)=0$ while the
element with $k=0$ satisfies precisely condition
\eqref{gapcondition}. In fact,  condition \eqref{gapcondition} is
the transformation of $P'(0)=0$ (which defines $\hat \cP_n$) under
the transformation \eqref{proj_action}.\qed

We would now want to know which operators  preserve the
transformed monomial subspace $\cM_n$. Of course if $T(z) \hat
\cP_n(z) \subset \hat \cP_n(z)$ then
$$
\tilde T(w) = (\gamma w+ \delta)^n\cdot T\left({\alpha w + \beta
\over \gamma w+ \delta}  \right)\cdot(\gamma w+ \delta)^{-n}
$$
will satisfy $\tilde T(w) \cM_n(w) \subset \cM_n(w)$. Transforming
the operators of Proposition \ref{prop5} under this prescription
will lead to a basis of $\fD_2(\cM_n)$. However, a simpler basis
can be written in the form expressed by the following proposition:
\begin{proposition}
A basis of $\fD_2(\cM_n)$ is given by the following seven
operators:
\begin{equation}\label{Jops}
\begin{aligned}
 J_1(w)  &= (\alpha w+\beta)^4\partial_{ww} + 2\alpha(1-n)(\alpha w+\beta)^3\partial_w + n(n-1)\alpha^2(\alpha w+\beta)^2\;,\\
 J_2(w)   &= (\alpha w+\beta)^3 \partial_{ww} +\alpha(1-n) (\alpha w+\beta)^2\partial_w\;,\\
J_3(w) &= (\alpha w+\beta)^2\partial_{ww} \;,\\
J_4(w)  & = (\alpha w+\beta)\partial_{ww} -\alpha (1+n\gamma (\alpha w+\beta))\partial_w \;,\\
J_5(w)   &=\partial_{ww}-{2\alpha\over(\alpha w+\beta)}\,\partial_w - 2n\alpha\gamma \partial_w + {2n \alpha^2\gamma \over (\alpha w+\beta)}\;,\\
J_6(w) &= (\alpha w+\beta)^2\partial_w - n \alpha (\alpha w+\beta) + \frac{(\alpha w+\beta)}{\gamma} \partial_w,\\
J_7(w) &= 1.
\end{aligned}
\end{equation}
\end{proposition}
\emph{Proof.} According to Proposition \ref{prop_basis}, a
different basis for $\cM_n(w)$ is given by
\begin{equation}\label{otherbasis}
 \{ y+ \frac{1}{n\gamma}, y^2,\dots,y^n
\;|\; y = \alpha w + \beta\},
\end{equation}
 and a straightforward computation
will verify that the set of operators
$$
\begin{aligned}
 & y^4\partial_{yy} + 2(1-n)y^3\partial_y + n(n-1)y^2\,,\\
 & y^3 \partial_{yy} +(1-n) y^2\partial_y\;,\\
 & y^2\partial_{yy}\,,\\
 & y\partial_{yy} -(1 + n\gamma y)\partial_y \;,\\
  & \partial_{yy}-2y^{-1}\partial_y- 2n\gamma \partial_y +{2n\gamma\over y}\;,\\
  & y^2\partial_y - n y + \frac{y}{\gamma} \partial_y,
\end{aligned}
$$
preserve the vector space spanned by \eqref{otherbasis}. Making
the appropriate substitutions yields directly \eqref{Jops}.\qed

%
%

\subsection{An example: the modified QES Morse potential}
\begin{figure}[htbp]
  \begin{center}
    \noindent\psfig{figure=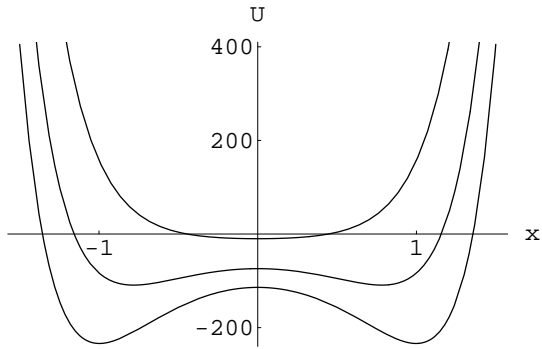,width=3in}
    \caption{ The deformed Morse potential $U_n(x)$ in \eqref{Ex:Morse} for $n=2$, $10$ and $15$.}
    \label{fig:morse}
  \end{center}
\end{figure}

Consider the following linear combination of the operators
\eqref{Jops}:
\begin{equation}\label{Tmorse}
T(z) = 8 J_3(z) - 8\sqrt{2} J_4(z) + 4 J_5(z) -16 J_6(z).
\end{equation}
If we choose $\alpha=\beta=\delta=1/{\sqrt{2}}$ and $\gamma =
-1/{\sqrt{2}}$ the above operator can be written in the form
\eqref{eq:Pzz} with
\begin{eqnarray}\label{PQR1}
&P(z)&= 4 z^2,\\
\label{PQR2}&Q(z) &= -4\left(2 z^2 + (n+1) z -4 +\frac{2}{1+z}  \right)\\
\label{PQR3} &R(z) &=8n(1+z) - \frac{4n}{1+z} .
\end{eqnarray}
The change of variables determined by \eqref{eq:physvar} is
$z=\e^{2x}$ and the potential is defined and free of
singularities for all values of $x$. In particular, the potential
form determined by \eqref{eq:potential} with
(\ref{PQR1})-(\ref{PQR3}) is
\begin{equation}\label{Ex:Morse}
U_n(x)=\frac{1}{8} \left(  \cosh{4x} - n \cosh{2x} +
\frac{1}{(1+\e^{2x})^2} - \frac{1}{1+\e^{2x}}\right)
\end{equation}
where an additive constant has been dropped. If the last two
terms were absent, this potential would be a QES deformation of
the Morse potential, which belongs to the classification of
Lie-algebraic potentials performed in \cite{GKO}. A plot of this
potential form for different values of $n$ can be found in
Figure~\ref{fig:morse}.
When $n$ is a non-zero integer, $n$ eigenfunctions of the
potential \eqref{Ex:Morse} can be calculated algebraically. They
have the following form:
\begin{equation}\label{factorised}
\psi_{k-1}(x)= \mu(x) \,p_k(\e^{2x}),\quad k=1,\dots,n
\end{equation}
where $\mu(x)$ is determined by \eqref{eq:gaugeform} with
(\ref{PQR1})-(\ref{PQR3}) to be
\begin{equation}
\mu(x)=\frac{\e^{-(\e^{-2x}+\e^{2x}+(n-1)x)}}{1+\e^{2x}}\,
\end{equation}
and $p_k(z)$ is one of the $n$ polynomial eigenfunctions that the
operator \eqref{Tmorse} has in the space
\begin{equation}\label{Mbasis2}
\mathcal M_n(z) = \myspan\{ z+ \frac{n-2}{n},(z+1)^2,\dots,(z+1)^n
\}.
\end{equation}
To be more specific, let us calculate a few eigenfunctions
explicitly. If we let $n=3$, the action of $T(z)$ with respect to
the basis \eqref{Mbasis2} of $\mathcal M_3(z)$ is given by:
\begin{equation}
T(z)|_{\mathcal M_3(z)}=\begin{pmatrix}
  \;-12\; & \;6\sqrt{2} \;&\; 0\; \\
  \;16\sqrt{2}\; & \;16\; & \;0\; \\
  \;0\; & \;8\sqrt{2}\; & \;36\;
\end{pmatrix}
\end{equation}
and the three algebraic eigenfunctions have been plotted in Figure
3.
\begin{figure}[htbp]
  \begin{center}
    \noindent\psfig{figure=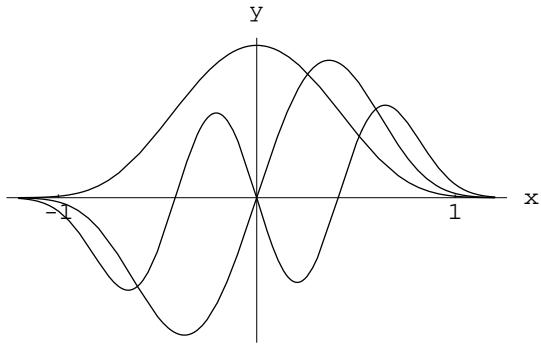,width=3in}
    \caption{ The three algebraic eigenfunctions $\psi_0$, $\psi_1$ and $\psi_2$ of $U_3(x)$.}
    \label{fig:morse_eigenf}
  \end{center}
\end{figure}
It is immediately apparent that the three algebraic functions
have $0$, $1$ and $3$ nodes, and therefore one eigenfunction of
the potential (the one with two nodes) is missing from the
algebraic sector, i.e. it cannot have the factorized form
\eqref{factorised}. The situation is repeated for higher values
of $n$ and our calculations suggest that in the general case, the
eigenfunction with $n-1$ nodes does not belong to the algebraic
sector. It remains an open question to find out what form this
``missing'' eigenfunction might have. This is the first example
of a quasi-exactly solvable potential in which the algebraic
eigenfunctions are not the lowest lying in the energy spectrum.

\section{Summary and conclusions}

We have shown how a more direct approach can be used to construct
the most general differential operator of a given order $r$ that
leaves invariant a polynomial subspace generated by monomials.
This direct approach can be used to address other problems that
would be very difficult to treat in the Lie-algebraic approach,
such as finding the most general differential operator that will
map a polynomial module $\cP_n$ into a proper subset
$\cP_{n-k}\subset \cP_n$. Problems of this kind will be treated
in our forthcoming publication \cite{GKM1}, where this approach
is also extended to deal with multivariate polynomials and
non-linear differential operators. The next step in difficulty
involves extending the analysis to non-monomial modules (i.e.
those which do not have a basis of monomials), a problem that
remains still open. The applications of the direct approach to
quasi-exact solvability in Quantum Mechanics point in two
directions. First, they provide a very natural explanation of why
some hamiltonians posses multiple algebraic sectors. As an
example, we discuss a generalization of the Lam\'e potential
\eqref{potLame}, which could have interesting applications in
solid state physics. Second, we emphasize that some quasi-exactly
solvable potentials exist which are not Lie-algebraic, as
illustrated by the modified QES Morse potential \eqref{Ex:Morse}.
In this example there is always one bound state missing from the
algebraic sector, which constitutes a new phenomenon in the
theory of quasi-exactly solvable problems.

\subsection*{acknowledgements}

The research of DGU is supported in part by a CRM-ISM
Postdoctoral Fellowship and the Spanish Ministry of Education
under grant EX2002-0176. The research of NK and RM is supported by
the National Science and Engineering Research Council of Canada.
DGU would like to thank the Department of Mathematics and
Statistics of Dalhousie University for their warm hospitality.

\end{document}